\documentclass[pre,showpacs, twocolumn, amsmath,amssymb]{revtex4}
\usepackage{psfig}
\begin{document}

\title{Understanding the internet topology evolution dynamics}

\author{Shi Zhou}\affiliation{University College London\\
Adastral Park Campus, Ross Building\\Ipswich, IP5 3RE, United Kingdom\\Email:
s.zhou@adastral.ucl.ac.uk}

\date{\today}

\begin{abstract}
The internet structure is extremely complex. The Positive-Feedback
Preference (PFP) model is a recently introduced internet topology
generator. The model uses two generic algorithms to replicate the
evolution dynamics observed on the internet historic data. The
phenomenological model was originally designed to match only two
topology properties of the internet, i.e.~the rich-club
connectivity and the exact form of degree distribution. Whereas
numerical evaluation has shown that the PFP model accurately
reproduces a large set of other nontrivial characteristics as
well. This paper aims to investigate why and how this generative
model captures so many diverse properties of the internet. Based
on comprehensive simulation results, the paper presents a detailed
analysis on the exact origin of each of the topology properties
produced by the model. This work reveals how network evolution
mechanisms control the obtained topology properties and it also
provides insights on correlations between various structural
characteristics of complex networks.

\end{abstract}


\maketitle

\section{Introduction}

It is vital to obtain a good description of a network topology
because structure fundamentally affects
function~\cite{albert02,pastor04}. The internet contains millions
of routers, which are grouped into thousands of subnetworks,
called \emph{autonomous systems} (AS), which are then glued into a
global network by the Border Gateway Protocol. Effective
engineering of the internet is predicated on a detailed
understanding of issues such as the large-scale structure of its
underlying physical topology, the manner in which it evolves over
time, and the way in which its constituent components contribute
to its overall function~\cite{floyd03}.

The recently introduced Positive-Feedback Preference (PFP)
model~\cite{zhou04d} is an internet AS-level topology generator.
The model uses two evolution mechanisms, namely the interactive
growth and the positive-feedback preference. Both mechanisms are
inspired by, and coincident with, practical observations on the
internet historic data. Originally the phenomenological model was
designed to match only the internet's rich-club
connectivity~\cite{zhou04a} and the exact form of degree
distribution, including the distribution of low degrees, the
maximum degree and the degree distribution's power-law
exponent~\cite{faloutsos99}. It has been a pleasant surprise that
numerical evaluation against the internet measurement date has
shown that the model accurately reproduces a large set of other
nontrivial topology characteristics as well, including the
disassortative mixing~\cite{newman02,vazquez03,maslov04}, the
shortest path length~\cite{watts98}, short
cycles~\cite{bianconi03a, caldarelli04} and the betweenness
centrality~\cite{goh01}. The PFP model has been used in more
realistic simulations of the internet, e.g.~\cite{nsf04-540}.

As of this writing, an analytical solution of the PFP model is not
available yet. In this paper we use the numerical method to
analyze why and how the model is able to reproduce a fuller
picture of the internet than it was designed for. In other words,
we aim to investigate the exact origin of each of the topology
properties captured by the PFP model. Answers to these questions
can be valuable for the ongoing effort on a mathematical solution
of the model.

In Section II, we review the PFP model and its two mechanisms. We
reflect on the intuitions underlying the design of the two
evolution mechanisms. In Section III, we comparatively examine two
limiting cases of the PFP model. Based on detailed numerical
simulations, we identify the explicit evolution dynamics that are
responsible for generating the rich-club phenomenon and the degree
distribution properties.  We also reveal that the origin of the
disassortative mixing is in fact already embedded in the model's
two evolution mechanisms. Furthermore we explain that the PFP
model resembles the internet's shortest path length and short
cycles because these two characteristics are correlated with other
topology properties.

In Section IV, the above results become more evident when we
examine how the PFP model's topology properties react to the
change of the parameters that control the model's evolution
mechanisms. Our investigation leads to a number of insightful
discoveries. For example we find out that the rich-club
connectivity is almost exclusively determined by the interactive
growth mechanism. We also show that the interactive growth and the
positive-feedback preference jointly contribute to the model's
disassortative mixing behavior, however the two mechanisms have
opposite effects on the degree distribution's power-law exponent.

In Section V we conclude that this work represents a fuller and
deeper understanding of the internet topology evolution dynamics.
This work complements the research on evolution mechanisms and
structural constrains of complex networks in general.

\section{The Positive-Feedback Preference Model}

In graph theory, degree,~$k$, is defined as the number of links a
node has. Degree is a local property but the distribution of
degree provides a global view of a network structure. The degree
distribution of the internet AS-level topology is characterized by
a power-law~\cite{faloutsos99} as
\begin{equation}
P(k) \sim k^{\gamma},~~\gamma\simeq-2.2. \end{equation} A
power-law degree distribution means the majority of nodes that
form a network have very few links. Some nodes, however, are very
well connected.

\begin{figure}[tbh]
\centerline{\psfig{figure=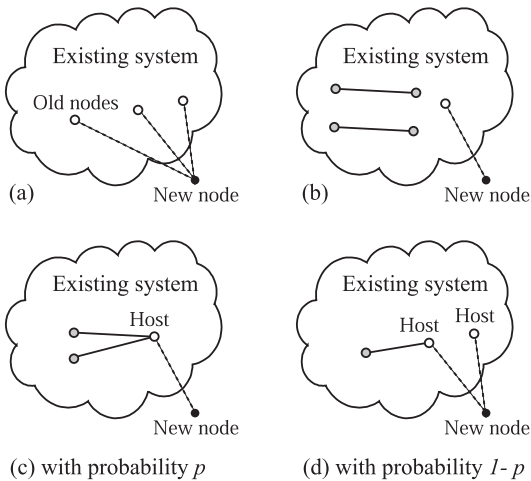,width=7cm}}
\caption{\label{fig:Growth} Network growth mechanisms: (a)~the BA
model's new-node-only growth; (b)~the independent growth; and (c)
and (d)~the PFP model's interactive growth.}
\end{figure}

Barab\'asi and Albert (BA)~\cite{barabasi99a} showed that a
power-law degree distribution can arise from two generic
mechanisms: \emph{growth}, where a network ``grows'' from a small
random network by attaching new nodes to $m$ old nodes in the
existing system (see Fig.\,\ref{fig:Growth}a, in which $m=3$); and
\emph{preferential attachment}, where a new node is attached
preferentially to nodes that are already well connected. The
probability that a new node will be connected to a node with
degree $k$ is given as
\begin{equation}
\Pi(k) = {k\over \sum_j k_j}. \label{eq:BA}
\end{equation}

During the last few years a large number of models have been
proposed to resemble and explain the power-law degree distribution
and other topology properties observed on various real
networks~\cite{albert02, pastor04}. Recently {Zhou} and
{Mondrag\'on}~\cite{zhou04d} introduced the Positive-Feedback
Preference (PFP) model. Numerical evaluation has shown that the
PFP model accurately reproduces a large number of characteristics
of the internet AS-level topology. The model uses the following
two evolution mechanisms.

\subsection{Mechanism One: Interactive
Growth}\label{section:IGmodel}

The interactive growth is designed to satisfy a number of
observations on internet history data\,\cite{vazquez02, chen02,
park03}. Firstly the internet evolution is largely due to two
processes: the attachment of new nodes to the existing system, and
the addition of new internal links between old nodes already
present in the existing system. Secondly the majority of new nodes
appeared on the internet are each attached to no more than two old
nodes. And thirdly the ratio of links to nodes on the internet AS
graph is approximately three.

Different from the independent growth~\cite{bu02} (see
Fig.\,\ref{fig:Growth}b), in which new nodes and new internal
links are added independently, the two evolution processes are
interdependent in the interactive growth (see
Fig.\,\ref{fig:Growth}c and d). That is to say, a new internal
link always starts from an old node, we call it a \emph{host}
node, to whom a new node has just attached. A heuristic
explanation of this interaction is that on the internet, new
customers (new nodes) generate extra demand for service, which
triggers their service providers (host nodes) to develop new
connections  to other service providers (new internal links) in
order to accommodate the increased traffic and improve services.

The interactive growth is described as such: starting from a small
random graph, at each time step,
\begin{itemize}
\item with probability $p\in[0, 1]$, a new node is attached to a
\emph{host} node, and at the same time two new internal links are
added connecting the host node to two other old nodes (see
Fig.\,\ref{fig:Growth}c); \item with probability $1-p$, a new node
is attached to two host nodes, and only one new internal link is
added connecting one of the host nodes to another old node (see
Fig.\,\ref{fig:Growth}d).
\end{itemize}
Numerical simulation shows that when the probability parameter
$p=0.4$, the interactive growth produces the best result.

\subsection{Mechanism Two: Positive-Feedback Preference}

The PFP model uses the following nonlinear preference probability
to choose old nodes for the interactive growth,
\begin{equation} \Pi(k)={k^{1+\delta\ln{k}}\over\sum_j
k_j^{1+\delta\ln{k_j}}},~~\delta\ge 0. \label{eq:PFP}
\end{equation}
It is called the positive-feedback preference (PFP) because a
node's ability of acquiring new links increases as a feed-back
loop of the node's degree. When the parameter $\delta=0$, it is
equivalent to the BA model's linear preference (see
Eq.\,\ref{eq:BA}). Numerical simulation shows that the
positive-feedback preference produces the best result when the
parameter $\delta=0.021$~\footnote{Different from \cite{zhou04d},
in this paper we use the natural logarithm in the
Eq.\,\ref{eq:PFP} of the positive-feedback preference. Accordingly
the value of parameter $\delta$ is revised as~$\delta_{new} =
{\delta_{old}/\ln{10}}$.}.

\subsection{Discussion}\label{section:pfp-discussion}

\begin{figure}[tbh]
\centerline{\psfig{figure=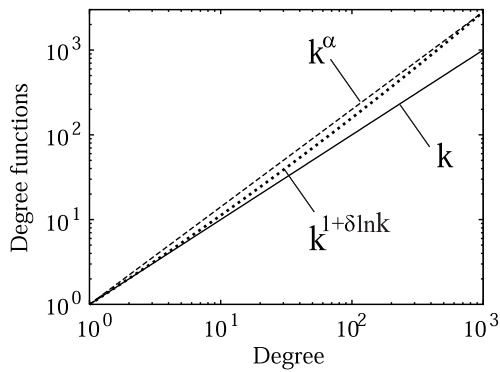,width=6.5cm}}
\caption{\label{fig:DegreeFunctions}Three degree functions:
$f(k)=k$ for the linear preference, $f(k)=k^{1.15}$ for the
exponential preference and $f(k)=k^{1+0.021\ln{k}}$ for the
positive-feedback preference.}
\end{figure}

The positive-feedback preference is designed to satisfy the
observation~\cite{park03, vazquez02, chen02} that during the
internet evolution, the probability that a new node links with a
low-degree node follows the BA model's linear preference, whereas
high-degree nodes have a stronger ability of acquiring new links
than predicted by the linear preference.

A preference probability can be written as
$\Pi(k)=f(k)/\sum_jf(k_j)$, where $f(k)$ is a degree function. The
exponential preference~\cite{dorogovtsev00,krapivsky00}, which
uses the nonlinear degree function of $f(k)=k^\lambda$,
$\lambda\geq1$, also gives high-degree nodes a stronger preference
than the linear preference. However as shown in
Fig.\,\ref{fig:DegreeFunctions}, only the PFP can be approximated
by the linear preference for low-degree nodes. When degree
increases, the PFP grows more and more rapidly and finally exceeds
the exponential preference.

\begin{table}[tbh] \centering
\caption{Preference probability ratio $f(\mu k)/f(k)$ when
${\lambda=1.15}$ and ${\delta=0.021}$.} \label{table:preference}
\begin{tabular}{c| c c c}
\hline\hline  &  Linear  & Exponential & Positive-Feedback \\
\hline
$f(k)$ & $k$  & $k^{\lambda}$  & $k^{1+\delta\ln{k}}$  \\
 $f(\mu k)\over f(k)$ & $\mu$ & $\mu^\lambda$ &
 $\mu^{1+\delta\ln{(\mu k)}}k^{\delta\ln{\mu}}$\\
\hline
$f(1000)\over f(100)$ & ${1000\over100}=10$ & ${2818.4\over199.5}=14.1$ & ${2722.7\over 156.1}=17.4$\\
${f(10)\over f(1)}$ &  ${10\over 1}=10$ & ${14.1\over 1}=14.1$ & ${11.2\over 1}=11.2$ \\
\hline\hline
\end{tabular}
\end{table}

\begin{figure}[tbh]
\centerline{\psfig{figure=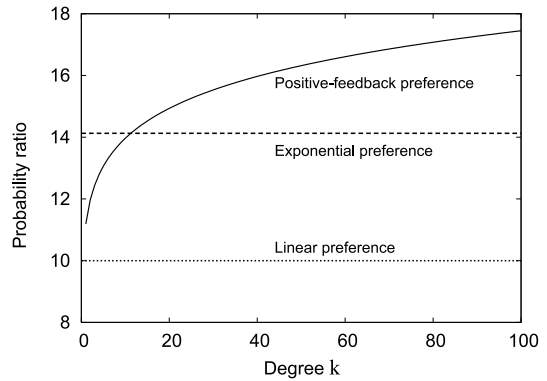,width=7cm}}
\caption{\label{fig:Ratio} Preference probability ratio $f(\mu
k)/f(k)$ as a function of degree $k$ when $\mu=10$, $\lambda=1.15$
and $\delta=0.021$. }
\end{figure}

To illustrate the impact of such difference, we compare a
$k$-degree node against a $\mu k$-degree node, $\mu\geqslant 1$.
The ratio of their preference probability can be given as $\Pi(\mu
k)/\Pi(k)=f(\mu k)/f(k)$. As shown in
Table\,\ref{table:preference}, for the linear and the exponential
preferences,  a {1000-degree} node, when competing against a
100-degree node, has the same advantage as a {10-degree} node
competing against a {1-degree} node. Whereas for the PFP,
$f(1000)/f(100)$ is more than 50\% larger than $f(10)/f(1)$. This
means the PFP not only appreciates the degree gap between
low-degree nodes and high-degree nodes, but also effectively
enlarges the degree difference between high-degree nodes
themselves (see Fig.\,\ref{fig:Ratio}). The consequence of the
positive-feedback preference is ``{rich not only get richer, they
get disproportionately richer}''.

\section{Evolution Mechanisms vs Topology Properties}

\begin{table*}[tbh]
\caption{Properties of the four models and the internet AS graph}
\label{table:fourmodels}\centering
\begin{tabular}{c|cccc|c}
\hline\hline
~&             BA  & IG  & BA+PFP  &  PFP  & AS graph\\
\hline
Growth mechanism     & { New-node-only}    & IG$_{~p=0.4}$ & {New-node-only}  &  IG$_{~p=0.4}$ &  \\
Preference scheme    & {Linear} & {Linear }& {PFP}$_{~\delta=0.021}$   & {PFP}$_{~\delta=0.021}$  &  \\
Number of nodes, $N$ & 9204& 9204& 9204& 9204&  9204 \\
Number of links, $L$ & 27612& 27612& 27612& 27612& 28959 \\
\hline
Rich-club phenomenon& weak & strong & weak & strong & strong\\
Rich-club exponent, $\theta$   &-0.97&-1.49& -0.97 &-1.49 & -1.481\\
Rich-club connectivity $\varphi(0.01)$   & 6.0\%  &40.3\%& 5.7\%  &  44.8\% &44.3\%\\
Top clique size, $n_{clique}$  &   2&  15& 4& 16 & 16\\
\hline
Degree distribution $P(1)$ &  0\% & 25.8\% & 0\% & 26.2\% &26.5\%\\
Degree distribution $P(2)$  &  0\%  & 33.7\% & 0\% &  33.7\%&30.2\% \\
Degree distribution $P(3)$  &  39.8\%  & 10.5\% & 43.7\% &  10.5\%&14.8\% \\
Degree distribution exponent, $\gamma$   &-2.902&-2.206& -2.890   & -2.255 & -2.254\\
Maximum degree, $k_{max}$       &240&677&898&1950 & 2070\\
\hline
Disassortative mixing& neutral & weak & weak &strong&strong\\
Assortativity coefficient, $\alpha$    &-0.036&-0.124&-0.091&-0.234& -0.236 \\
\hline
Characteristic path length, $\ell^*$ &4.25&3.55&3.75&3.07&3.12\\
\hline\hline\end{tabular}
\end{table*}

To investigate the relations between the PFP model's evolution
mechanisms and the obtained topology properties, we compare the
PFP model and the BA model against the following two limiting
cases.
\begin{enumerate}
\item The Interactive Growth (IG) model, which uses the PFP
model's interactive growth and the BA model's linear preference.
\item The BA+PFP model, which uses the BA model's new-node-only
growth and the PFP model's positive-feedback preference.
\end{enumerate}
For each of the four models, we generate ten networks to the same
size as the internet AS graph~\cite{itdk0403,mahadevan05a} using
different random seeds. All networks are generated from small
random graphs consisting of 10 nodes randomly connected by 30
links. Quantities in Table~\ref{table:fourmodels} are averages
over the ten networks. Detailed evaluation of the PFP model
against Internet measurement data has been provided
in~\cite{zhou04d}. In the following we focus on the comparison of
topology properties among the four models.

\subsection{Rich-Club Phenomenon}\label{section-richclub}

A hierarchical structure of the internet AS-level topology is the
rich-club phenomenon~\cite{zhou04a}, which describes the fact that
well connected nodes, \emph{rich} nodes, tent to be tightly
interconnected with other rich nodes forming a core group, or
club. Rich-club membership can be defined as ``the $r$ best
connected nodes'', where $r$ is a node's {rank} denoting the
node's position on the non-increasing degree list of a network.
Node rank is often normalized by $N$,  the number of nodes
contained in the network.

The rich-club phenomenon is assessed by measuring the rich-club
connectivity, $\varphi(r/N)$, defined as the ratio of the actual
number of links to the maximum possible number of links among the
rich-club members. The internet AS graph is fundamentally
characterized by a power law of $\phi(r/N)\sim r^{\theta}$ with
the exponent $\theta=-1.48$, which results from fitting
$\phi(r/N)$ with a power law for 90\% of the nodes,
i.e.~$0.1\leqslant r/N \leqslant1$. Rich-club connectivity
indicates how well club members ``know'' each other,
e.g.~$\varphi=1$ means that all the members have a direct link to
any other member, i.e. they form a fully connected mesh, a clique.
The tope clique size, $n_{_{clique}}$, defined as the maximum
number of nodes with the highest ranks still forming a clique.

\begin{figure}[tbh]
\centerline{\psfig{figure=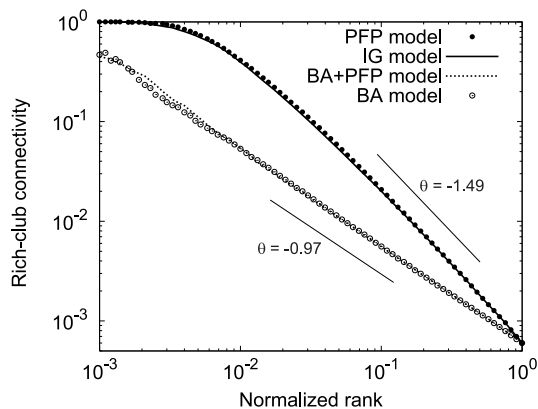,width=7cm}}
\caption{Rich-club connectivity vs normalized rank,
$\varphi(r/N)$.} \label{fig:Rich}
\end{figure}

As shown in Fig.\,\ref{fig:Rich} and Table~\ref{table:fourmodels},
the PFP model and the IG model exhibit the same rich-club
connectivity. So do the BA model and the BA+PFP model. However the
former two models, using the interactive growth, produce a
significantly stronger rich-club phenomenon  than that obtained by
the later two models using the new-node-only growth. It is evident
that the rich-club phenomenon is primarily determined by the
growth mechanisms, not the preference schemes.

The BA model and the BA+PFP model use the new-node-only growth
mechanism, in which all newly added links are external links
between new nodes and old nodes. The  old nodes are preferentially
chosen from rich nodes but the new nodes have only three initial
links. Thus the only chance for the rich-club connectivity to
increase is when new nodes become rich nodes, however  this
usually happens in the early stage of network growth, and later on
new nodes become more and more difficult to compete for new links
against the already rich nodes. As a result, although rich nodes
keep acquiring new links, the interconnections \emph{among} rich
nodes hardly increases.

On contrast the PFP model and the IG model use the interactive
growth mechanism, which adds not only external links, but also
internal links between old nodes. Since the old nodes are
preferentially chosen from already rich nodes, the newly added
internal links directly increase the rich-club connectivity. The
ratio of internal links to external links can be estimated  as a
function of the parameter~$p$,
\begin{equation}
{L_{int}\over L_{ext}} = {2p+(1-p) \over p+ 2(1-p)} =
{1+p\over2-p}.\label{eq:link-ratio}
\end{equation}
Simulation result shows that when~$p=0.4$,
i.e.~$L_{int}/L_{ext}=7/8$, the interactive growth produces a
rich-club phenomenon that precisely matches that of the internet
AS graph.

The reason that using different preference schemes has little
impact on the obtained rich-club phenomenon is that, although
preference schemes influence the degree growth rate (i.e.~how many
links rich nodes have), they have little effect on the
interconnectivity among the rich nodes.

\subsection{Degree Distribution Properties}

\begin{figure}[tbh]
\centerline{\psfig{figure=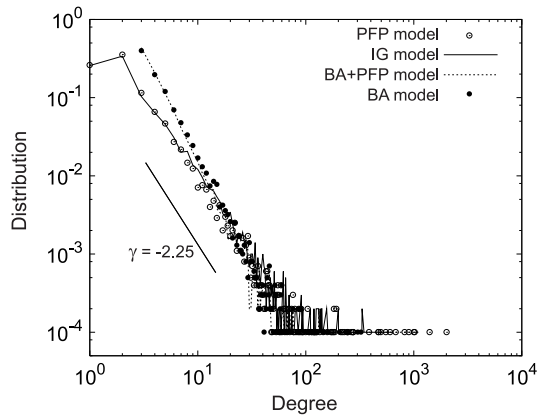,width=7cm}} \caption{Degree
distribution, $P(k)$.} \label{fig:PK}
\end{figure}

\subsubsection{Distribution of Low Degrees}

Table~\ref{table:fourmodels} and Fig.\,\ref{fig:PK} show that more
than half of the nodes on the internet AS graph are 1-degree or
2-degree nodes, and the distribution of low degrees does not
follow a power law because $P(k=1)=26.5\%<P(k=2)=30.2\%$. The BA
model and the BA+PFP model do not contain 1-degree and 2-degree
nodes because they use the new-node-only growth, in which each new
node is attached to three old nodes, i.e.~$k\geqslant3$. The IG
model and the PFP model use the interactive growth mechanism,
which assigns a new node's initial degree as one or two according
to the probability parameter~$p$. Simulation result shows when
$p=0.4$ the interactive growth closely resembles the AS graph's
distribution of low degrees as well.

\subsubsection{The Maximum Degree and Degree Growth Rate}

As shown in Fig.\,\ref{fig:PK}, the AS graph's degree distribution
does not follow a strict power law because it has a heavy tail.
The maximum degree, $k_{max}$, is the largest degree in a network.
The maximum degree is an indicator of how far the degree
distribution deviates from the prediction of a strict power law.
Table~\ref{table:fourmodels} shows that the internet AS graph
features a very large maximum degree, which is one order of
magnitude larger than that generated by the BA model.

\begin{figure}[tbh]
\centerline{\psfig{figure=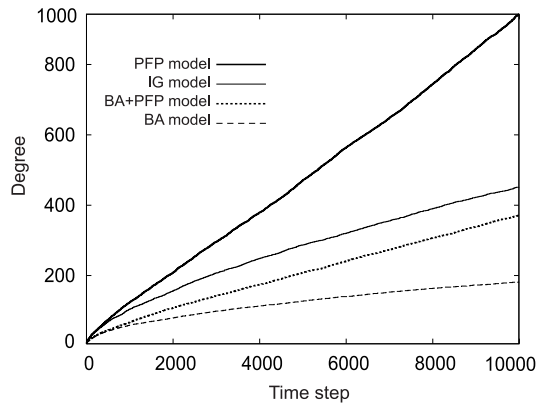,width=7cm}} \caption{
Average degree growth of the initial nodes. } \label{fig:KT}
\end{figure}

Fig.\,\ref{fig:KT} illustrates the growth of average degree  of
nodes contained in the initial small random graphs from which the
networks grow. These nodes are the earliest ones present in the
networks and usually they represent a sample of the rich nodes in
the generated networks. As expected, the nodes in the BA+PFP model
enjoy a higher degree growth rate than the BA model because the
positive-feedback preference gives stronger preference to
high-degree nodes than the BA model's linear preference.

Fig.\,\ref{fig:KT} also shows that the degree growth rate of the
IG model is higher than the BA model and the BA+PFP model. This
means the interactive growth is also able to accelerate the link
acquiring process. As shown in Fig.\,\ref{fig:Growth}, the
new-node-only growth and the independent growth only allow a
chosen old node to acquire one new link per time step. Whereas the
interactive growth is fundamentally different. It enables a chosen
old node, i.e.~a host node, to acquire up to three new links at
each time step: with probability $p$, a host node acquires one
external link and two internal links; and with probability $1-p$,
a host node acquires one external link and one internal link. We
call this the \emph{degree-leap} effect of the interactive growth.
The degree-leap effect increases the degree of host nodes by two
or three, which would take many time steps to achieve when using
other growth mechanisms. The degree-leap effect also significantly
enhances the node's ability to compete for even more new links in
the forthcoming time steps, therefore the node's degree growth
rate is  accelerated.

The PFP model uses both the interactive growth and the
positive-feedback preference, which reinforce each other, to
achieve a degree growth rate that is notably more rapid than any
other models. As a result, the PFP model obtains a very large
maximum degree, as large as that of the internet AS graph.

\subsubsection{Power-Law Exponent of Degree Distribution}

As shown in Fig.\,\ref{fig:PK}, all four models produce power-law
degree distributions, but have different power-law exponents,
which result from fitting $P(k)$ in the area of $2\leqslant k
\leqslant100$. Table~\ref{table:fourmodels} shows that the
power-law exponents produced by the IG model and the PFP model,
both using the interactive growth, are close to that of the
internet AS graph, $\gamma=-2.254$. Whereas the exponent generated
by the BA model and the BA+PFP model is close to
$\gamma=-3.0$~\cite{albert02}.

It is clear that the interactive growth has a major impact on the
obtained degree distribution exponent. This is because the
interactive growth has the degree-leap effect,  which allows more
nodes to be ``fast-tracked'' into rich nodes and also makes the
already rich nodes get richer more rapidly. The consequences of
this dynamics is that the obtained power-law degree distribution
exhibits a flatter slope (with a larger value of $\gamma$) and a
heavier tail (with a larger maximum degree).

Studying the degree distribution exponents shown in
Table~\ref{table:fourmodels}, we can see that the exponent of the
IG model, $\gamma=-2.206$, is actually overly increased by the
interactive growth and slightly larger than that of the internet
AS graph. The PFP model accurately matches the AS graph's exponent
because its positive-feedback preference has a minor effect of
reducing the value of $\gamma$. This is because, comparing with
the linear preference, the PFP gives a strong favor to the richest
nodes (at the tail of the degree distribution) at the cost of all
other nodes. As a result, the degree distribution is slightly
steeper (with a smaller value of $\gamma$), the power law cuts off
at a smaller degree, and the tail gets even longer.

\subsection{Disassortative Mixing}

Networks exhibit different degree-degree mixing
patterns~\cite{newman02, vazquez03, maslov04}. For example social
networks are usually classified as assortative networks, because
nodes in social networks statistically tend to attach to alike
nodes, i.e.~high-degree nodes to high-degree nodes and low-degree
nodes to low-degree nodes. On contrast, technological and
biological networks, e.g.~the internet, exhibit the disassortative
mixing, where high-degree nodes tend to connect with low-degree
nodes, and visa versa. The BA model, however, is a neutral network
which exhibits no mixing tendency.

The rich-club phenomenon observed on the internet does not
conflict with the fact that the internet is a disassortative
network, because the rich-club phenomenon does not imply that the
majority links of the rich nodes are directed to other rich-club
members. Indeed, rich nodes have very large numbers of links and
only a few of them are enough to provide the interconnectivity to
other club members, whose number is anyway small~\cite{pastor04}.

\begin{figure}[tbh]
\centerline{\psfig{figure=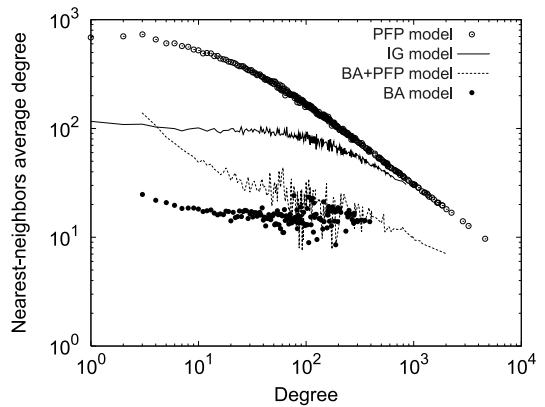,width=7cm}}
\caption{Nearest-neighbors average degree vs node degree.}
\label{fig:Knn}
\end{figure}

A network's mixing pattern is decided by the conditional
probability $p_c(k'|k)$ that a link connects a $k$-degree node to
a $k'$-degree node. For computational simplicity, a network's
mixing pattern is often identified by the correlation between node
degree and nearest-neighbors average degree~\cite{vazquez02}. As
shown in Fig.\,\ref{fig:Knn}, the PFP model is a disassortative
network because it exhibits a negative degree-degree correlation,
and the BA model is a neutral network because the
nearest-neighbors average degree is almost invariant to the node
degree.

Another way to determine a network's mixing pattern is a metric
called the assortativity coefficient~\cite{newman02},
$\alpha\in[-1, 1]$, which is defined as
\begin{equation}
\alpha = {L^{-1}\sum_{i} j_i k_i -
[L^{-1}\sum_i{1\over2}(j_i+k_i)]^2 \over
L^{-1}\sum_i{1\over2}(j_i^2+k_i^2)-[L^{-1}\sum_i{1\over2}(j_i+k_i)]^2},
\end{equation}
where $L$ is the number of links, and $j_i$, $k_i$ are the degrees
of the nodes at the ends of the $i$th link, with $i=1,2,...,L$.
When $\alpha>0$, a network is an assortative network, and when
$\alpha<0$, it is a disassortative network. As shown in
Table~\ref{table:fourmodels}, the internet AS graph, characterized
by a negative assortativity coefficient of $\alpha=-0.236$, is
closely resembled by the PFP model. The assortativity coefficient
of the BA model is close to zero. In between are that of the IG
model and the BA+PFP model.

This result shows that both the positive-feedback preference and
the interactive growth contribute to a network's disassortative
mixing behavior. As we have discussed in
Section~\ref{section:pfp-discussion}, the positive-feedback
preference can effectively amplify  the degree difference between
two nodes. Thus any degree difference between a link's two end
nodes will be magnified by the PFP because the end node with a
larger degree will grow faster and faster than the other end node.
As a result the PFP increases a network's disassortative mixing.
The interactive growth also increases the disassortative mixing
but in a different way. When external links are attached between
new nodes and host nodes, the host nodes are preferentially chosen
from already rich nodes, and they enjoy the extra degree growth
given by the interactive growth's degree-leap effect. However the
new nodes are to remain as low-degree, poor nodes. Thus the
interactive growth introduces external links with a larger degree
gap between the end nodes than other growth mechanisms do.

As shown in the IG model and the BA+PFP model, when either the
interactive growth or the positive-feedback preference works
alone, their effect on strengthening  the disassortative mixing is
limited. When the two mechanisms work together, as shown in the
PFP model, they reinforce each other and generate networks as
disassortative as the internet AS graph.

Up to this point, we have identified the explicit evolution
dynamics that are responsible for the above topology properties.
We can see the origin of these topology properties are embedded in
the PFP model's two evolution mechanisms.

\subsection{Shortest Path Length}

The internet is a \emph{small-world} network~\cite{watts98}
because it is possible to get to any node via only a few links
among adjoining nodes. The shortest path length, $\ell$, is the
minimum hop distance between a pair of nodes. Performance of
modern network routing algorithms depends strongly on the
distribution of shortest path length~\cite{labovits01}. The
characteristic path length, $\ell^*$, is the average shortest path
length over all pairs of nodes.

The characteristic path length of the internet AS graph is only
3.12 hops. The internet is so small is because it exhibits both a
strong rich-club phenomenon and a strong disassortative mixing.
These two structural properties together contribute to the routing
efficiency of the network.  The rich-club consists of a small
number of highly connected nodes. The club members are tightly
interconnected between each other. If two club members do not have
a direct connection, they are very likely to share a neighboring
club member. Thus the average hop distance between the members is
very small (between 1 and 2 hops). The rich-club functions as a
``super'' traffic hub of a network by providing a large selection
of shortcuts for routing. The disassortative mixing property
ensures that the majority of network nodes, peripheral low-degree
nodes, are always near the rich-club. Thus a typical shortest path
between two peripheral nodes consists of three hops, the first hop
is from the source node to a member of the rich-club, the second
hop is between two club members and the final hop is to the
destination node.

\begin{figure}[tbh]
\centerline{\psfig{figure=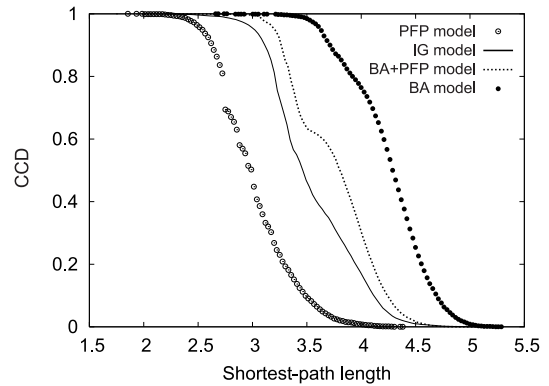,width=7cm}}
\caption{Complimentary cumulative distribution (CCD) of
shortest-path length, $P(\geqslant\ell)$.} \label{fig:PL}
\end{figure}

As shown in Fig.\,\ref{fig:PL} and Table~\ref{table:fourmodels},
the PFP model accurately reproduces the rich-club phenomenon and
the disassortative mixing of the internet AS graph, naturally it
reproduces the internet's shortest path length as well. The BA
model exhibits a weak rich-club phenomenon and it is a neutral
network. As a result the BA model's characteristic path length is
more than one hop longer than that of the internet AS graph. This
one-hop difference is significant considering that the AS graph's
characteristic path length is merely over three hops. The BA+PFP
model exhibit a rich-club phenomenon as weak as the BA model,  but
it exhibits a (weak) disassortative mixing. Consequently its
characteristic path length is half-hop shorter than that of the BA
model. The IG model is more disassortative than the BA+PFP model
and it exhibits the rich-club phenomenon as strong as the PFP
model, therefore the IG model is smaller than the BA+PFP model,
but not as small as the PFP model.

\subsection{Short Cycles -- Triangle Coefficient}

Short cycles, i.e.~triangles and quadrangles, encode the
redundancy information in a network because the multiplicity of
paths between any two nodes increases with the density of short
cycles~\cite{caldarelli04,bianconi03a}. The triangle coefficient,
$k_t$, is defined as the number of triangles that a node shares,
or the number of links connecting among a node's nearest
neighbors. The clustering coefficient~\cite{watts98} can be given
as $c={k_t\over k(k-1)/2}$. Comparing with the clustering
coefficient, the triangle coefficient is able to infer neighbor
clustering information of nodes with different degrees, i.e.~the
correlation between triangle coefficient and degree.

\begin{figure}[tbh]
\centerline{\psfig{figure=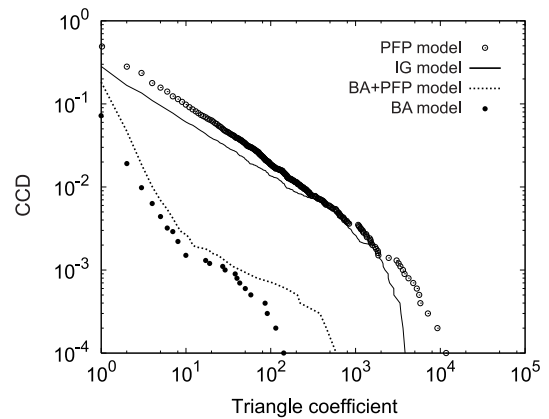,width=7cm}}
\caption{Complement cumulative distribution of node triangle
coefficient, $P(\ge k_t)$.}\label{fig:P-Tri}
\end{figure}

\begin{figure}[tbh]
\centerline{\psfig{figure=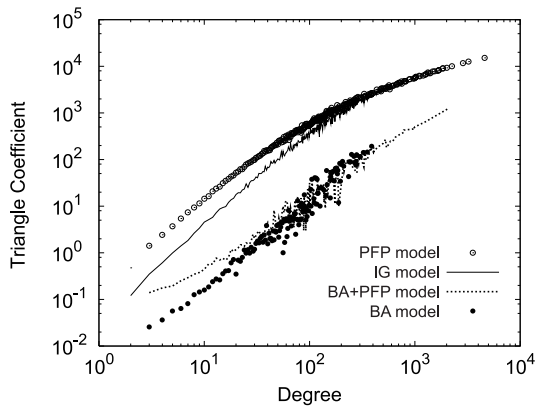,width=7cm}}
\caption{Average triangle coefficient of $k$-degree nodes,
$k_{t}(k)$.}\label{fig:K-Tri}
\end{figure}

Fig.\,\ref{fig:P-Tri} shows that the density of triangles of the
PFP model and the IG model is significantly larger than that of
the BA model and the BA+PFP model. Fig.\,\ref{fig:K-Tri} shows
that the average triangle coefficient of $k$-degree nodes in the
PFP model and the IG model is one order of magnitude larger than
that of nodes with the same degrees in the BA model and the BA+PFP
model. It is evident that models showing a stronger rich-club
phenomenon contain considerably more triangles. This is because
the interconnections between rich-club members play a major role
on the formation of triangles, not only for the club members
themselves, but also for the peripheral low-degree nodes which
have more than one connections to the rich-club.

The disassortative mixing also has a minor impact on the formation
of triangles. When a model exhibits a stronger disassortative
mixing, links of the low-degree nodes are more likely connected to
the rich-club, and thus form more triangles. For example, the PFP
model and the IG model show the same rich-club phenomenon but the
PFP model exhibits a stronger disassortative mixing. Thus, as
shown in Fig.\,\ref{fig:K-Tri}, the average triangle coefficient
of nodes in the degree area of $2<k<100$ in the PFP model is
notably larger than that of nodes with the same degrees in the IG
model. The same can be observed by contrasting the BA+PFP model
against the BA model in the degree area of $3<k<30$.

In summary, the PFP model captures the internet's shortest path
length and triangle coefficient because these properties are
correlated with other topology properties that have been produced
by the model's evolution mechanisms.

\section{Sensitivity To Model Parameters}

The PFP model's interactive growth and positive-feedback
preference are controlled by the parameter $p$ and $\delta$
respectively. In the previous section the PFP model uses $p=0.4$
and $\delta=0.021$ to generate internet-like networks. In this
section we provide more detailed numerical results to support the
above analysis.  We examine how the parameters control the
evolution mechanisms and therefore change the generated topology
properties. We first study the model's sensitivity to individual
parameters by making one parameter a variable and fixing the other
as a constant. Then we investigate the model's reactions when both
parameters are variables.

\subsection{Sensitivity To Parameter $p$}

\begin{table*}
\caption{The PFP model's sensitivity to parameter $p$ when
$\delta=0$.} \label{table:parameter-p} \centering
\begin{tabular}{c| c c c c c |c}
\hline\hline Interactive growth parameter $p$&    $0.0$  & $0.2$ &
$0.4$ & $0.6$ & $0.8$&AS graph\\
\hline $L_{int}/ L_{ext}={(1+p)/(2-p)}$ & 1/2 & 2/3 & 7/8 & 8/7 & 3/2 &\\
Rich-club exponent, $\theta$    & -1.36 &-1.43&  -1.49  &-1.56& -1.61 & -1.481 \\
Rich-club connectivity $\varphi(0.01)$   & 26.1\%  &33.3\%& 40.3\%  &48.2\%&  53.8\% &44.3\%\\
Top clique size, $n_{clique}$ &   7 &10& 15 &17& 20 & 16\\
\hline
Degree distribution $P(1)$ & 0\%  &13.1\%&  25.8\% &38.9\%&  50.5\%  &26.5\%\\
Degree distribution $P(2)$  & 49.5\%  &41.6\%& 33.7\%  &25.9\%&  18.7\% &30.2\% \\
Degree distribution $P(3)$  & 13.5\%  &12.1\%& 10.5\%  &8.4\%&  6.6\% &14.8\% \\
Degree distribution exponent, $\gamma$  & -2.416  &-2.229&-2.206  &-2.151& -2.055&-2.254 \\
Maximum degree, $k_{max}$      &  573  &625& 677&722& 763  &2070\\
\hline
Assortativity coefficient, $\alpha$    & -0.075  &-0.095& -0.124&-0.150& -0.183  &-0.236\\
Characteristic path length, $\ell^*$     & 3.65& 3.60&3.55 & 3.52  & 3.48&3.12\\
\hline\hline\end{tabular}
\end{table*}

\begin{figure}[tbh] \vspace{5mm}
\centerline{\psfig{figure=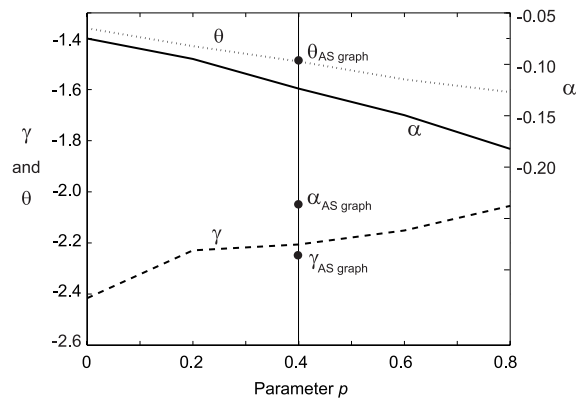,width=7.5cm}}
\caption{Properties of the PFP model when parameter $p$ grows from
0 to 0.8 while $\delta=0$, including the degree distribution
exponent $\gamma$, the rich-club connectivity exponent $\theta$,
and the assortativity coefficient $\alpha$.}
\label{fig:parameter-p}
\end{figure}

Table~\ref{table:parameter-p} and Fig.\,\ref{fig:parameter-p} show
how topology properties of the PFP model change when the
interactive growth parameter $p$ increases from $0$ to $0.8$ while
the positive-feedback parameter $\delta=0$ (i.e.~equavelant to the
linear preference). It is clear that when $p$ increases, the
rich-club phenomenon is getting stronger and stronger as the value
of rich-club exponent $\theta$ decreases monotonically. This is
because the ratio of new internal links to new external links
added to the model increases with $p$ (see
Eq.\,\ref{eq:link-ratio}). When more internal links are added, the
rich nodes become more tightly interconnected. Also the
interactive growth has a direct impact on the distribution of low
degrees. When $p$ increases, the generated network contains more
1-degree nodes and less 2-degree nodes (see $P(1)$ and $P(2)$ in
Table~\ref{table:parameter-p}). When $p=0.4$, the interactive
growth well resembles the internet AS graph's rich-club phenomenon
and distribution of low degrees at the same time.

As analyzed in the above, the interactive growth has a degree-leap
effect. When $p$ increases, more 3-degree leaps (see
Fig.~\ref{fig:Growth}c) and less 2-degree leaps (see
Fig.~\ref{fig:Growth}d) take place during the network growth. Thus
the overall degree-leap effect become stronger. As a result the
power-law degree distribution becomes flatter with an increased
value of the degree distribution exponent~$\gamma$. As shown in
Fig.\,\ref{fig:parameter-p}, when $p$ increases to 0.4,
$\gamma$~exceeds that of the internet AS graph. We will see in the
next section that this excessive increase of~$\gamma$ will be
counter balanced by the positive-feedback preference's opposite
effect on~$\gamma$.

When $p$ grows from $0$ to $0.8$, the strengthened degree-leap
effect also enlarges the obtained maximum degree, and strengthens
the model's disassortative mixing behavior indicated by a
decreasing value of the assortativity coefficient. As expected,
when the rich-club phenomenon and the disassortative mixing become
stronger, the generated networks get smaller and smaller indicated
by a decreasing value of the characteristic path length. However,
when $p=0.8$ and the model exhibits a stronger rich-club
phenomenon than the internet AS graph, the PFP model's
characteristic path length is still not as small as the internet.
This is because the model's disassortative mixing is yet as strong
as the Internet. In order to resemble the internet, the
interactive growth needs to be combined with the positive-feedback
preference.

\subsection{Sensitivity To Parameter $\delta$}

\begin{table*}
\caption{The PFP model's sensitivity to parameter $\delta$ when
$p=0.4$.} \label{table:parameter-delta} \centering
\begin{tabular}{c| c c c c c c|c}
\hline\hline Positive-feedback parameter $\delta$
& $0$ & $0.007$ & $0.014$ & $0.021$ & $0.028$ & $0.035$&AS graph\\
\hline
Rich-club exponent, $\theta$  & -1.49 & -1.49&-1.49     & -1.49& -1.50 &-1.44& -1.481 \\
Rich-club connectivity $\varphi(0.01)$   & 40.3\%  &44.0\%& 43.5\%  &44.8\%&  46.4\% &36.5\%&44.3\%\\
Top clique size, $n_{clique}$  &  15 &16 &16  &16 &20  &15& 16\\
\hline
Degree distribution $P(1)$ &  26.2\% & 26.4\%& 26.8\%& 26.2\%&   26.4\%&22.0\%&26.5\%\\
Degree distribution $P(2)$  &  33.7\%  & 34.3\%& 34.8\%& 33.7\%&   35.2\%&34.7\%&30.2\% \\
Degree distribution $P(3)$  &  10.5\%  & 10.6\%& 10.5\%& 10.5\%&   12.4\%&16.7\%&14.8\% \\
Degree distribution exponent, $\gamma$     & -2.206 & -2.219& -2.228  & -2.255&  -2.321&-2.540&-2.254 \\
Maximum degree, $k_{max}$       &  677  & 875& 1356& 1950& 2519&7045& 2070\\
\hline
Assortativity coefficient, $\alpha$     & -0.124  & -0.172& -0.202& -0.234&  -0.279&-0.292&-0.236\\
Characteristic path length, $\ell^*$    & 3.55  &3.41 & 3.24&3.07 &  2.93  & 2.44 & 3.12\\
\hline\hline\end{tabular}
\end{table*}

\begin{figure}[tbh]
\centerline{\psfig{figure=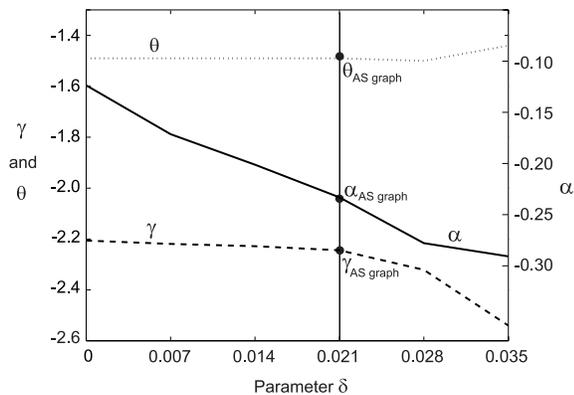,width=7.5cm}}
\caption{Properties of the PFP model when parameter $\delta$ grows
from 0 to 0.035 while $p=0.4$, including the degree distribution
exponent $\gamma$, the rich-club connectivity exponent $\theta$,
and the assortativity coefficient
$\alpha$.}\label{fig:parameter-delta}
\end{figure}

Table~\ref{table:parameter-delta} and
Fig.\,\ref{fig:parameter-delta} show how topology properties of
the PFP model change  when the positive-feedback parameter
$\delta$ increases from 0 to 0.035 while the interactive growth
parameter $p=0.4$. It is clear that the positive-feedback
parameter $\delta$ has a fairly limited impact on the obtained
rich-club exponent $\theta$, which almost remains the same. When
$\delta$ increases, greater preference is given to high-degree
nodes, consequently the maximum degree increases and the network
becomes more disassortative mixed. As expected, the characteristic
path length decreases steadily when the network's disassortative
mixing is getting stronger. When the parameter $\delta$ increases,
the degree distribution power-law exponent $\gamma$ slightly
decreases. This is because when the positive-feedback preference
gets stronger, the richest nodes attract so many new links that
they suppress the degree growth of other nodes, including those
with medium to high degrees. When $\delta=0.021$ (and $p=0.4$),
the PFP model closely matches all the topology properties of the
internet AS graph.

As shown in Fig.\,\ref{fig:parameter-delta}, by tuning the
parameter $\delta$ from 0 to 0.028, the PFP model is capable of
generating a wide range of disassortative mixing networks with the
value of the assortativity coefficient $\alpha$ decreasing
monotonically from -0.12 to -0.28, which encompasses most
technological and biological networks reported in~\cite{newman03}.
Notably, the PFP model achieves this while keeping the rich-club
exponent and the degree distribution exponent largely unchanged.

$\delta=0.028$ is the model's tipping point, beyond which the
network structure changes dramatically, e.g.~the degree
distribution exponent $\gamma$ decreases rapidly. This is because
the preferential selection becomes so biased that super-rich nodes
start to emerge. These extremely rich nodes dominate the network
growth and make the network grow towards a star-like structure.

\subsection{Sensitivity To Both Parameters}

\begin{figure}[tbh]
\centerline{\psfig{figure=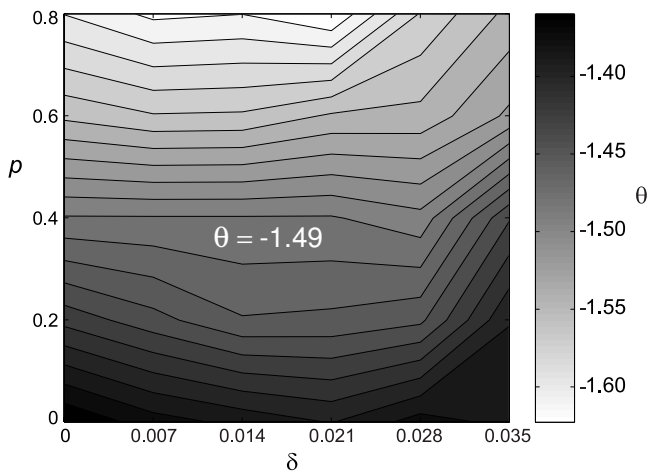,width=8.5cm}}
\caption{Rich-club connectivity's power-law exponent $\theta$ vs
parameters~$p$ and~$\delta$.} \label{fig:parameter-theta}
\end{figure}

\begin{figure}[tbh]
\centerline{\psfig{figure=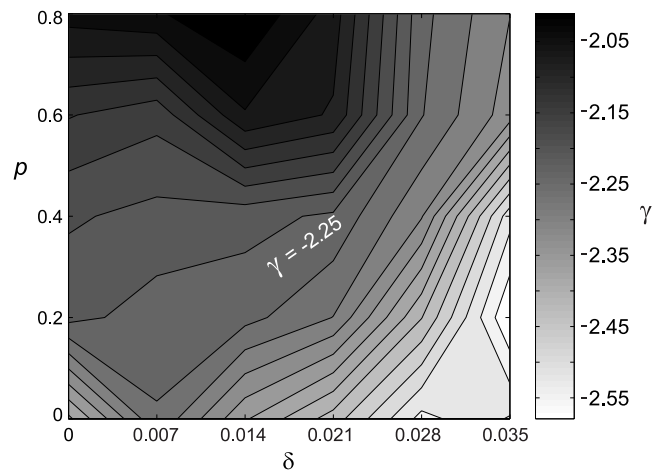,width=8.5cm}}
\caption{Degree distribution's power-law exponent $\gamma$ vs
parameters~$p$ and $\delta$.} \label{fig:parameter-gamma}
\end{figure}

\begin{figure}[tbh]
\centerline{\psfig{figure=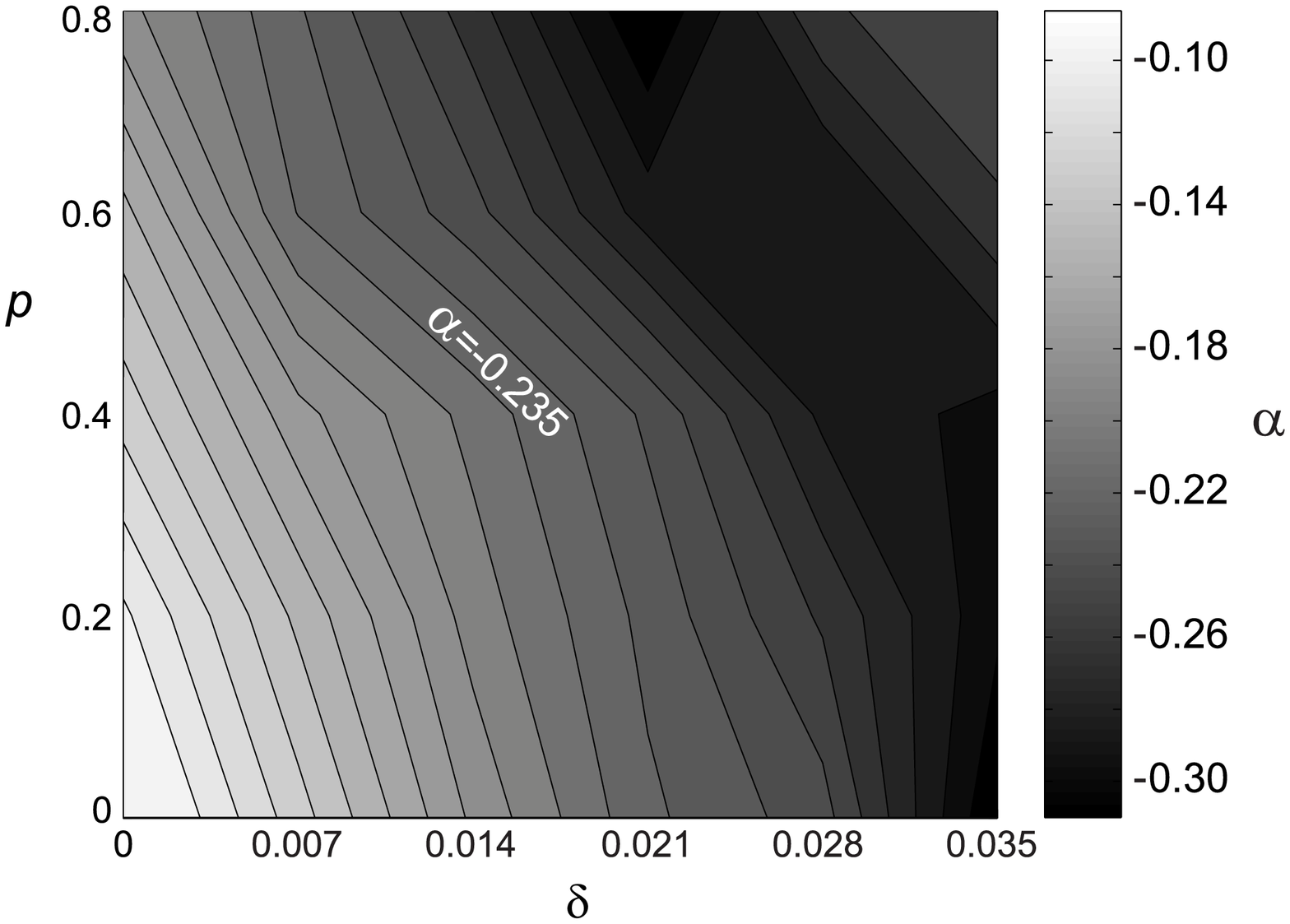,width=8.5cm}}
\caption{Assortativity coefficient $\alpha$ vs parameters~$p$ and
$\delta$.} \label{fig:parameter-alpha}
\end{figure}

Fig.\,\ref{fig:parameter-theta}--\ref{fig:parameter-alpha} are
contour plots showing how three topology properties, i.e.~the
rich-club connectivity exponent $\theta$, the degree distribution
exponent $\gamma$ and the assortativity coefficient $\alpha$,
change when both the parameters are variables.
Fig.\,\ref{fig:parameter-theta} clearly shows that the value of
rich-club exponent $\theta$ is sensitive to the interactive growth
parameter $p$ and is unsensitive to the positive-feedback
preference parameter $\delta$. When parameter $p$ grows, the
rich-club phenomenon becomes stronger.
Fig.\,\ref{fig:parameter-gamma} shows that the two mechanisms have
opposite effects on the value of degree distribution exponent
$\gamma$. In general, the exponent $\gamma$ increases when
parameter $p$ increases and parameter $\delta$ decreases.

Fig.\,\ref{fig:parameter-alpha} shows that the PFP model's
disassortative mixing becomes stronger  when either of the two
parameters increases. The assortativity coefficient $\alpha$ is
more sensitive to parameter $\delta$ than to parameter~$p$. We
notice that when $p>0.5$, the network's disassortative mixing
actually becomes weaker. This is because, when $p>0.5$, the
network starts to acquire more internal links than external links,
i.e.~$L_{int}/ L_{ext}\ge 1$ (see Eq.~\ref{eq:link-ratio}). The
increased new internal links among the rich nodes weaken the
disassortative mixing behavior of the network.

\section{Conclusion}

The internet has an extremely complex structure. The PFP model
demonstrates a way to simultaneously reproduce many topology
properties of the internet. The model achieves this by using two
generic algorithms which are designed to replicate the evolution
dynamics observed on the internet historic data. In this paper we
have used the numerical method to investigate the success of the
model.

Based on detailed simulation results, we show that the rich-club
phenomenon is primarily controlled by the interactive growth
mechanism alone. We point out that this is because the rich-club
connectivity increases with the number of new internal links which
are added between rich nodes. The interactive growth also
determines the probability of new nodes' initial degrees and thus
controls the distribution of low degrees obtained in the generated
network.

The PFP model's maximum degree is as large as the internet because
both the interactive growth and the positive-feedback preference
accelerate the degree growth rate. The interactive growth does so
by the degree-leap effect, whereas the positive-feedback
preference achieves so by giving stronger preference to
high-degree nodes. The two mechanisms have opposite effects on the
value of power-law exponent of degree distribution. The
interactive growth increases the value whereas the
positive-feedback preference slightly decreases the value.

We have explained that the origin of the disassortative mixing has
been, unintentionally, embedded in the PFP model's two evolution
mechanisms, which not only enlarge the gap between the high-degree
nodes and the low-degree nodes, but more critically, they increase
the degree difference between end nodes of a link.

Our analysis suggests that the reason that the PFP model also
captures other properties of the internet, such as  shortest path
length and triangle coefficient, is that these properties are
correlated with other properties, in other words, they are
constrained by other. For example the internet is small because
the rich-club functions as a super traffic hub, while the
disassortative mixing behavior ensures peripheral low-degree nodes
always close to the hub.

By investigating the PFP model's sensitivity to the mechanism
parameters, we obtain more evident details on how the two
evolution mechanisms profoundly control the model's growth
dynamics and therefore effectively change the generated topology
properties. For example we show that by tuning the
positive-feedback parameter $\delta$, the PFP model is able to
generate a wide range of disassortative mixing networks while
keeping  the rich-club exponent and the degree distribution
exponent almost unchanged.

This work represents a better understanding of the internet
topology evolution dynamics. It facilitates the ongoing effort on
a mathematical analysis of the PFP model. This work also
complements the research on the evolution dynamics and the
structural constrains of complex networks in general.

\end{document}